\documentclass{nature}

\usepackage{graphicx}
\usepackage{lineno}
\usepackage{amsmath}
\usepackage{color, soul}
\title{Magnetism and Berry Phase Manipulation in an Emergent Structure of Perovskite Ruthenate by (111) Strain Engineering}
\author{Zhaoqing Ding$^{1,2,\dagger}$, Xuejiao Chen$^{3,\dagger}$, Zhenzhen Wang$^{1,2,\dagger}$, Qinghua Zhang$^{1}$, Fang Yang$^{1}$, Jiachang Bi$^{3}$, Ting Lin$^{1,2}$, Zhen Wang$^{1,2,4}$, Xiaofeng Wu$^{1,2}$, Minghui Gu$^{1,2}$, Meng Meng$^{1}$, Yanwei Cao$^{3}$, Lin Gu$^{5}$, Jiandi Zhang$^{1}$, Zhicheng Zhong$^{3,6,\ast}$, Xiaoran Liu$^{1,\ast}$, Jiandong Guo$^{1,2,\ast}$
}

\begin{document}

\maketitle

\begin{affiliations}
 \item Beijing National Laboratory for Condensed Matter Physics and Institute of Physics, Chinese Academy of Sciences, Beijing 100190, China.
 \item School of Physical Sciences, University of Chinese Academy of Sciences, Beijing 100049, China.
 \item Ningbo Institute of Materials Technology and Engineering, Chinese Academy of Science, Ningbo, Zhejiang 315201, China.
 \item Institute of High Energy Physics, Chinese Academy of Sciences, Beijing 100049, China.
 \item Beijing National Center for Electron Microscopy and Laboratory of Advanced Materials, Department of Materials Science and Engineering, Tsinghua University, Beijing 100084, China.
 \item College of Materials Science and Opto-Electronic Technology, University of Chinese Academy of Sciences, Beijing 100049, China
\end{affiliations}

\noindent{$^\ast$ Correspondence should be addressed to X.L. (xiaoran.liu@iphy.ac.cn) or Z.Z. (zhong@nimte.ac.cn) or J.G. (jdguo@iphy.ac.cn)}

\newpage

\begin{abstract}

The interplay among symmetry of lattices, electronic correlations, and Berry phase of the Bloch states in solids has led to  fascinating quantum phases of matter. A prototypical system is the magnetic Weyl candidate SrRuO$_3$, where designing and creating electronic and topological properties on artificial lattice geometry is highly demanded yet remains elusive. Here, we establish an emergent trigonal structure of SrRuO$_3$ by means of heteroepitaxial strain engineering along the [111] crystallographic axis. Distinctive from bulk, the trigonal SrRuO$_3$ exhibits a peculiar XY-type ferromagnetic ground state, with the coexistence of high-mobility holes likely from linear Weyl bands and low-mobility electrons from normal quadratic bands as carriers. The presence of Weyl nodes are further corroborated by capturing intrinsic anomalous Hall effect, acting as momentum-space sources of Berry curvatures. The experimental observations are consistent with our first-principles calculations, shedding light on the detailed band topology of trigonal SrRuO$_3$ with multiple pairs of Weyl nodes near the Fermi level. Our findings signify the essence of magnetism and Berry phase manipulation via lattice design and pave the way towards unveiling nontrivial correlated topological phenomena. 

\end{abstract}

\newpage

%{\noindent \bf Introduction}\

The notion of Berry phase has been recognized a powerful and unifying concept in many different fields of physics since its discovery in 1984 \cite{Berry_1984_PRSA}. In condensed matter physics, application of this concept has brought out geometric and topological viewpoints, generating deeper understandings of various transport phenomena \cite{Xiao_RMP_2010}. Of particular interest is the intrinsic contribution to the anomalous Hall effect (AHE), which is determined by integrating the Berry curvature of each occupied Bloch state over the entire Brillouin zone \cite{Nagaosa_RMP_2010}. Since such a scattering-independent Hall conductivity only depends on the electronic band structure, the Berry-phase mechanism has established a link between the AHE and the topological nature of materials \cite{Narang_2020_NM}. Especially in ferromagnetic conductors with broken time-reversal symmetry and sizable spin-orbit couplings, peculiar features such as avoided crossings or Weyl points near the Fermi energy can act as the source and sink of the Berry curvatures, significantly enhancing the intrinsic contribution and dominating the AHE over other extrinsic mechanisms (i.e. side-jump and skew scattering) \cite{Nagaosa_RMP_2010,Armitage_2018_RMP,Nagaosa_2020_NRM,Burkov_2018_ARCMP,Schoop_2018_CM}.   

A prototypical system of this category is the perovskite ruthenate SrRuO$_3$. The dual nature of both itinerancy and localization of Ru 4$d$ electrons makes SrRuO$_3$ a rare example of ferromagnetic metallic oxides \cite{Koster_2012_RMP,Hahn_2021_PRL}. The nontrivial topological properties of SrRuO$_3$ were initially realized as a result of the non-monotonous relationship between its anomalous Hall conductivity $\sigma_{xy}$ and the magnetization $M$, which could be rationalized well from first-principle calculations by taking into account the Berry phases of electronic band structures \cite{Fang_2003_science}. These were subsequently suggested as the existence of Weyl nodes characteristic of linear band crossings at the Fermi energy \cite{Itoh_2016_NC,Takiguchi_2020_NC,Kaneta_2022_NPJQUANTMATS,Kar_2023_NPJQUANTMATS}. Lately, a variety of experimental efforts have been put forth to tailor the Berry-phase related features of (001)-oriented SrRuO$_3$ films via epitaxial strain \cite{Tian_2021_PNAS}, helium irradiation \cite{Skoropata_2021_PRB}, variation of film thickness \cite{Takada_2021_APL,Sohn_2021_NM,Ali_2022_NPJQUANTMATS}, ionic liquid gating \cite{Sohn_2021_NM}, and construction of asymmetric interfaces \cite{Groenendijk_2020_PRR}. In addition, it was argued in (001) SrRuO$_3$ ultrathin films and heterostructures that topologically nontrivial features can also emerge in real space due to the formation of nonlinear spin textures such as the skyrmions, giving rise to the so-called `topological Hall effect' (THE) \cite{Matsuno_2016_SA,Wang_2018_NM,Wang_2019_NM}. Nevertheless, the contribution of THE to the net Hall effect is usually entangled with those from the Berry curvatures in momentum space and practically challenging to be unambiguously identified \cite{Wang_2020_NL,Kimbell_2020_PRM,Wu_2020_PRB,Kimbell_2022_CM}.

\begin{figure}[htp]
\centering
\includegraphics[width=0.95\textwidth]{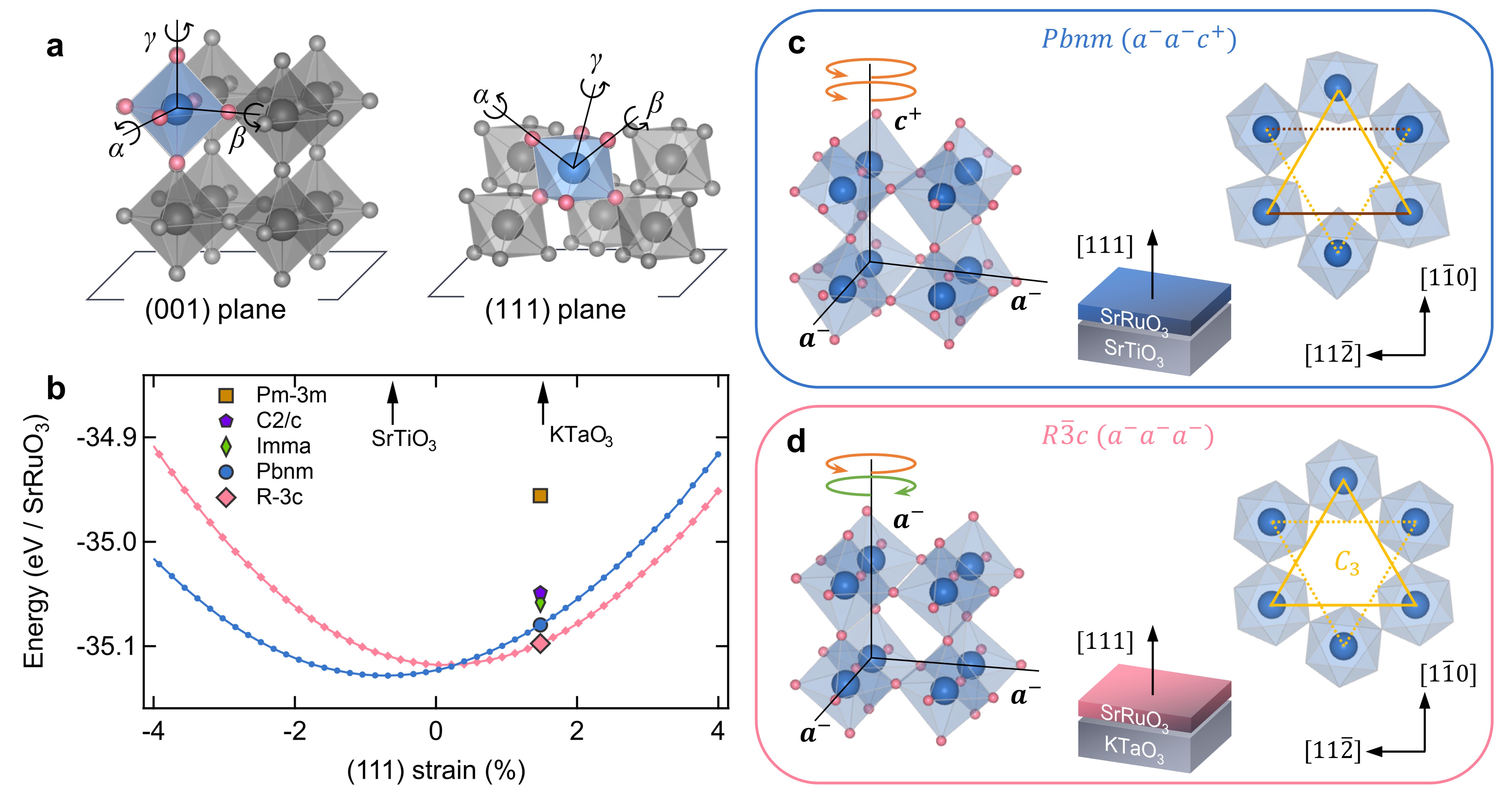}
\caption{\label{Fig1}
{\bf Emergence of hidden trigonal SrRuO$_3$ under (111) tensile strain.} \textbf{a,} Schematic depiction of the octahedral tilt and rotation axes for (001) and (111) orientations, respectively. All indices of the crystallographic planes and directions in this work are defined in the frame of pseudocubic axes unless for special notice. \textbf{b,} Energy of structures vs. magnitude of (111) strain by first-principles calculation. Six possible phases are considered on KTaO$_3$ substrate and the lowest five in energy are displayed on the figure. \textbf{c-d,} Heteroepitaxial stabilization of the orthorhombic $Pbnm$ and the trigonal $R\bar{3}c$ SrRuO$_3$ on (111) SrTiO$_3$ and KTaO$_3$ substrates, respectively. The rotation patterns are denoted in the Glazer notations \cite{Glazer_1972_AC}.}
\end{figure}

Recently, inspired by the salient studies on the honeycomb lattice \cite{Haldane_1988_PRL,Kane_2005_PRL}, heterostructures of perovskite-type transition metal oxides along the pseudocubic [111] axis have been extensively investigated by theorists as promising platforms towards exotic correlated and topological phenomena \cite{Xiao_2011_NC, Wang_2011_PRB, Ruegg_2011_PRB, Ruegg_2013_PRB,Wang_2015_PRB, Si_2017_PRL, Marthinsen_2018_PRM}. Notably, the symmetry of the (111) surface is trigonal and the three octahedral rotation axes lie neither parallel nor perpendicular to the (111) plane, as illustrated in Fig.~\ref{Fig1}a. As a result, the strategy of strain accommodation in (111) heteroepitaxy can manifest in rather distinct manners, which may essentially affect the topological phase diagrams and/or lead to emergent phenomena that are inaccessible in bulk and (001)-oriented heterostructures \cite{Ruegg_2013_PRB, Kim_2014_PRB, Moreau_2017_PRB,Chakhalian_2020_APLM,Cuoco_2022_APLM}. From the experimental perspective, while initiative studies were mainly focused on strongly correlated 3$d$ complex oxides \cite{Ueda_1998_Science, Gibert_2012_NM, Kim_2016_Nature, Middey_2016_PRL, Hepting_2018_NP, Asaba_2018_PRB, Kane_2021_npjQM}, it has remained largely unexplored in ruthenates with intermediate strength of correlations and sizable spin-orbit interactions \cite{Chang_2009_JCG,Rastogi_2019_APLM,Lin_2021_AM}.

Here, we report the discovery of an emergent trigonal phase of SrRuO$_3$ stabilized epitaxially on (111) KTaO$_3$ substrate. Unlike the original $a^-a^-c^+$ rotation pattern in bulk, the $\sim$1.5\% tensile strain triggers the formation of $a^-a^-a^-$ pattern in trigonal SrRuO$_3$. Combined magnetization and transport measurements have revealed its peculiar magnetic and topological properties: (1) three-dimensional (3D) XY ferromagnetism (FM) with sixfold anisotropic magnetoresistance (AMR) below 160 K, (2) coexistence of high-mobility holes ($\sim$10$^4$ cm$^2$V$^{-1}$s$^{-1}$) and low-mobility electrons as charge carriers, (3) double AHE channels including intrinsic contributions from the Berry-phase mechanism. These results are consistent with our first-principles electronic structure calculations, which suggest the presence of multiple pairs of Weyl nodes within an energy range of 50 meV at the Fermi level. Our findings highlight the opportunity for unveiling hidden correlated topological phenomena by means of (111) strain engineering. \\

%exhibiting electron pockets from normal quadratic bands and hole pockets from the linearly dispersive Weyl bands. 

{\noindent \bf Results and Discussion}\

{\noindent \bf Heteroepitaxial stabilization of $R\bar{3}c$ SrRuO$_3$ under (111) tensile strain.} 
Our earlier study about SrRuO$_3$ films on (111) SrTiO$_3$ substrate has uncovered that the compressive strain ($\sim$-0.6\%) is accommodated by significantly suppressing the degree of the $c^+$ octahedral rotation \cite{Wang_2021_CPB}. Thus, it is desirable to explore the strategy of strain accommodation and the feasibility of achieving latent phases of SrRuO$_3$ on the tensile strain side. Here, we select KTaO$_3$ as the substrate in that it possesses the same perovskite-type structure as SrRuO$_3$ and can supply a practically large magnitude of tensile strain of $\sim$1.5\%.

\begin{figure}[htp]
\centering
\includegraphics[width=0.9\textwidth]{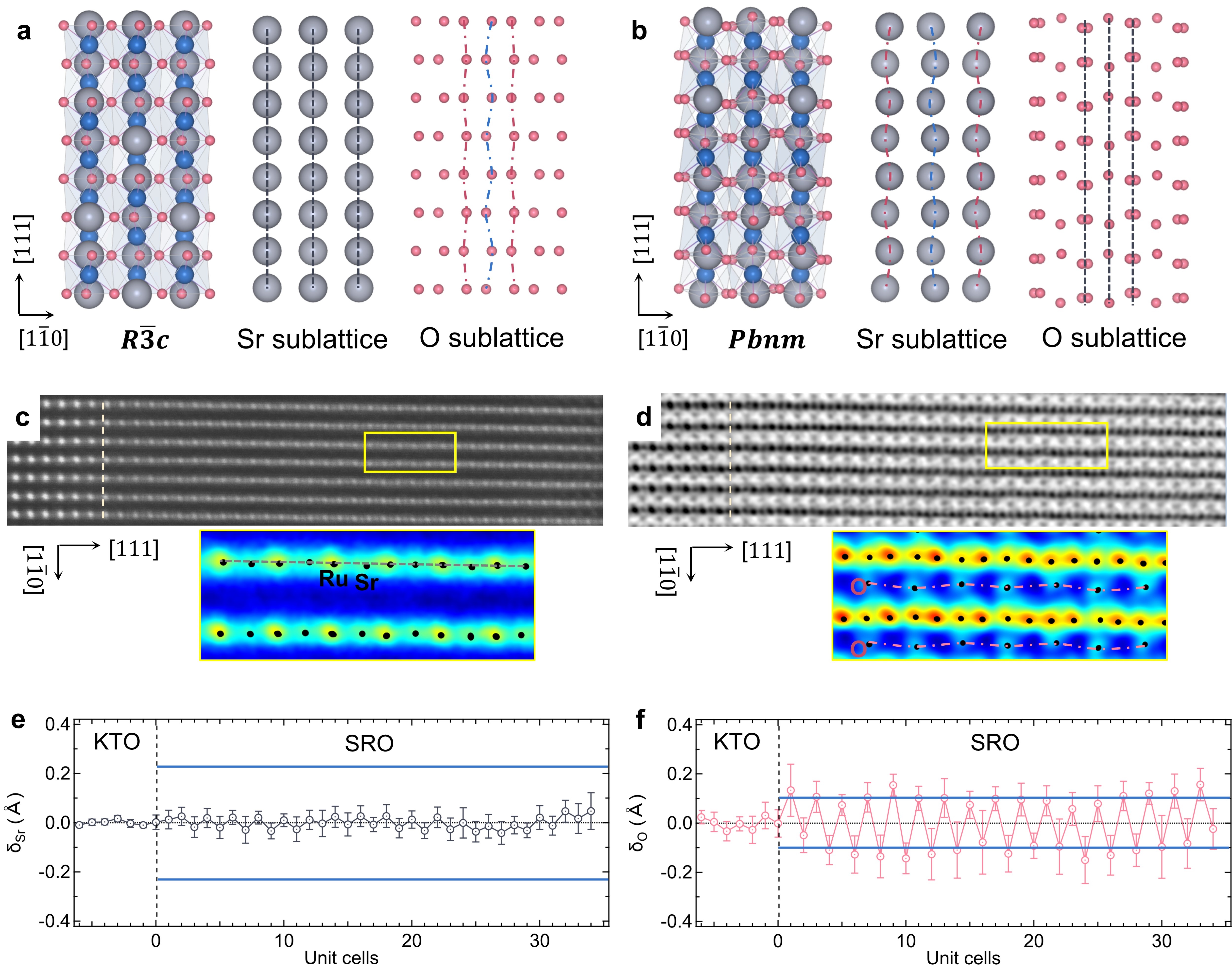}
\caption{\label{Fig2}
{\bf Atomic-scale illustration of the trigonal SrRuO$_3$ in (111) epitaxial films on KTaO$_3$ substrate.} \textbf{a-b,} Characteristic crystal structures of SrRuO$_3$ in \textbf{(a)} $R\bar{3}c$ and \textbf{(b)} $Pbnm$ space group projected along the  [11$\bar{2}$] direction. The Sr and the O sublattices are displayed separately. For each symmetry as a result of the corresponding octahedral tilt and rotation, the atoms with relative displacements are indicated by red and blue dash lines, while those of no relative displacements are indicated by black dash lines. \textbf{c,} HAADF and \textbf{d,} ABF STEM images of the heterostructure. The areas in SrRuO$_3$ enclosed by the yellow boxes on individual figures are enlarged below, where the levels of intensities are converted into different colors using the CalAtom software\cite{Zhang_2019_UltraMS}. The calculated positions of atoms are marked by black dots. \textbf{e,} Layer-resolved evolution of the atomic relative displacements within the Sr columns. Note, the positions of Ta atoms on KTaO$_3$ side are counted as references. The blue lines represent the ideal values in the $Pbnm$ phase. \textbf{f,} Layer-resolved evolution of the atomic relative displacements within the O columns across the interface. The blue lines represent the theoretical values in the $R\bar{3}c$ phase. In \textbf{e} and \textbf{f}, the standard error of each data point is achieved by statistically sampling along the vertical [1$\bar{1}$0] direction.}
\end{figure}

%Thickness of the (111) SrRuO$_3$ films is $\sim$15 nm as identified from the x-ray reflectivity (XRR) measurements (Fig.~\ref{Fig1}d). The symmetric x-ray diffraction (XRD) scans around the KTaO$_3$ (222) reflection (inset of Fig.~\ref{Fig1}d) reveal the 2$\theta$ position of SrRuO$_3$ (222) reflection at $\sim$87.313$^\circ$, leading to a putative pseudocubic lattice constant of $\sim$3.87 \AA. This value is rather smaller than the value of $\sim$3.93 \AA\ of bulk SrRuO$_3$ within the Pbnm space group, signifying the out-of-plane spacing is distinctly decreased plausibly due to tensile strain induced by the (111) KTaO$_3$ substrate (lattice constant $\sim$3.99 \AA). The reciprocal space mapping (RSM) around the KTaO$_3$ asymmetric (123) reflection further confirms this hypothesis (Fig.~\ref{Fig1}e). The SrRuO$_3$ (123) reflection is clearly observed possessing the same in-plane $Q_X$ as the substrate, indicating the SrRuO$_3$ thin film is coherently strained on (111) KTaO$_3$. 

We have synthesized (111)-oriented SrRuO$_3$ thin films ($\sim$15 nm) on KTaO$_3$ substrates using the pulsed laser deposition technique (see `Methods'). Scanning transmission electron microscopy (STEM) image confirms the expected epitaxial relationship between film and substrate with abrupt interface (Supplementary Figure 1). Comprehensive x-ray diffraction characterizations demonstrate that the films have high crystallinity without secondary phases, which are coherently strained on the substrate (Supplementary Figure 2).   

To obtain microscopic information about the octahedral rotation patterns, the schematic crystal structures of SrRuO$_3$ with $R\bar{3}c$ and $Pbnm$ space groups are projected along the [11$\bar{2}$] direction, respectively, highlighting their peculiar distinctions in atomic arrangements on the Sr and O sublattices (Fig.~\ref{Fig2}a and \ref{Fig2}b). Specifically, in the $R\bar{3}c$ phase, while the Sr atoms in each column along [111] are aligned straightforwardly (Fig.~\ref{Fig2}a, black dash lines), the O atoms exhibit pronounced displacements forming opposite zigzag patterns between adjacent columns and the same pattern in every other column (Fig.~\ref{Fig2}a, red and blue dash lines). In sharp contrast, in the $Pbnm$ phase, displacements of the Sr atoms exhibit the zigzag patterns whereas the O columns are aligned straightforwardly. Hence it can be regarded as the hallmarks for identifying the symmetry of our tensile-strained (111) SrRuO$_3$ thin films.

Indeed, these characteristic features are captured by high-precision STEM high-angle annular dark field (HAADF) and annular bright field (ABF) images, as exhibited in Fig.~\ref{Fig2}c and \ref{Fig2}d. It is evident that the Sr atoms are lined up, meanwhile, the relative displacements of the O atoms in every other column give rise to the same zigzag pattern, all compatible with the $R\bar{3}c$ characters. This is further corroborated by quantitative analyses on the layer-resolved atomic displacements across the interface. There are overall no experimentally resolvable Sr displacements, far from the ideal magnitude in the $Pbnm$ phase as indicated by the blue lines on Fig.~\ref{Fig2}e. On the other hand, the single O columns exhibit well-defined zigzag patterns on SrRuO$_3$ side, with the magnitudes of relative displacements in good agreement with the theoretical value (Fig.~\ref{Fig2}f). These results lead us to conclude that the trigonal SrRuO$_3$ phase that is inaccessible in bulk has been stabilized by means of (111) strain engineering.\\

\begin{figure}[htp]
\centering
\includegraphics[width=0.95\textwidth]{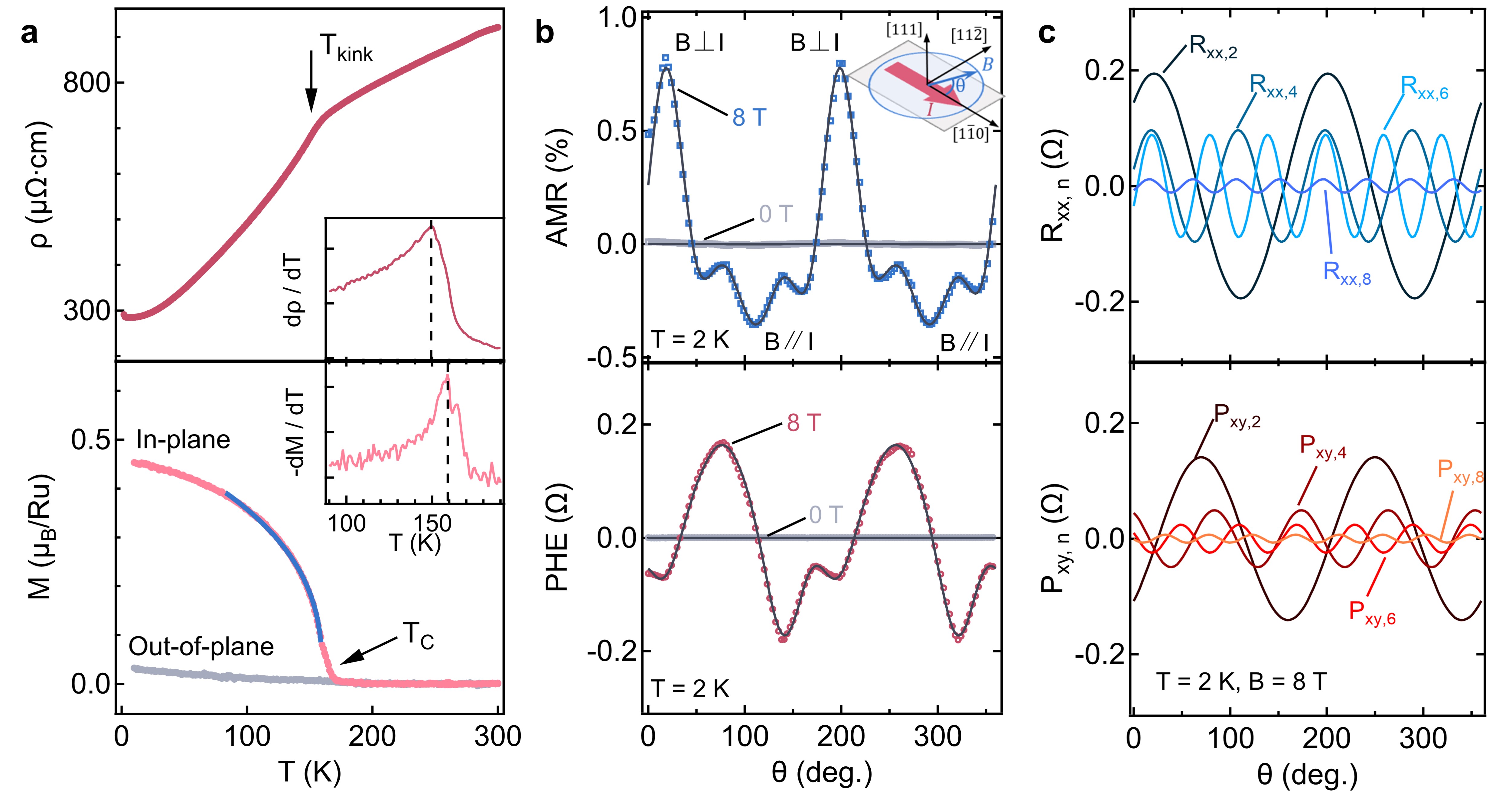}
\caption{\label{Fig3}
{\bf Electronic and magnetic properties of trigonal SrRuO$_3$.} \textbf{a,} Temperature dependence of resistivity and remnant magnetization. The inset exhibits the derivative curves highlighting individual transition temperature as labeled on the graph (T$_\textrm{kink}$ and T$_\textrm{C}$). The critical scaling fit on M(T) (blue solid line) close to the phase transition determines an exponent $\beta \sim$ 0.36, falling in the three-dimensional XY universality class \cite{Campostrini_2000_PRB}. \textbf{b,} Angular dependence of AMR and PHE measured in 8 T and 0.1 T magnetic field, respectively. Data are displayed in open square (circle) for AMR (PHE), and the black curves are fits from the equations given in the main text. The inset shows the geometry of measurements: the current was fixed along the [1$\bar{1}$0] direction and the magnetic field was applied within the (111) plane with an angle $\theta$ relative to the [1$\bar{1}$0] direction. The measurements started at an offset of $\theta$ around 67$^\circ$. \textbf{c,} The magnitude of each harmonics extracted from the corresponding fits.}
\end{figure}

{\noindent \bf Electronic and magnetic properties of trigonal SrRuO$_3$.} 
Next, we unravel the magnetic properties of trigonal SrRuO$_3$. The temperature dependence of resistivity $\rho$(T) indicates an overall metallic behavior down to 2 K, with a kink present at T$_\textrm{kink}$ $\sim$150 K reflecting the para- to ferro-magnetic transition (Fig.~\ref{Fig3}a, top panel). The onset of ferromagnetism is clearly visible on the temperature-dependent spontaneous magnetization $M$(T) curve with a Curie temperature T$_\textrm{C}$ $\sim$160 K (Fig.~\ref{Fig3}a, bottom panel). These values are in general close to those reported in bulk and (001)-oriented films \cite{Koster_2012_RMP}. However, unlike the uniaxial magnetic anisotropic behaviors observed in the $Pbnm$ phase \cite{Klein_1996_PRL}, it is noteworthy that the trigonal SrRuO$_3$ displays planar magnetic anisotropy below T$_\textrm{C}$, and the moments lie in the (111) plane (Supplementary Figure 3). A good fit to the in-plane spontaneous magnetization, $M(T) = M_0 (1-T/T_\textrm{C})^\beta$, gives rise to the critical exponent $\beta$ = 0.36, consistently falling in the three-dimensional (3D) XY universality class \cite{Campostrini_2000_PRB}.   

More insights into the magnetic state are obtained by measuring the {\it in-plane} anisotropic magnetoresistance (AMR) and planar Hall effect (PHE). In general, both AMR and PHE should exhibit dominant twofold oscillations as a function of $\theta$ in a full cycle with 45$^{\circ}$ relative phase shift (i.e. $R_{xx}$ $\sim$ $cos2\theta$; $R_{xy}$ $\sim$ $sin2\theta$), due to rotation of the principle axes of the resistivity tensor \cite{Huang_2021_PRR}. However, symmetry of the lattice and magnetic structure can induce additional anisotropy, giving rise to higher harmonics of oscillations (fourfold, sixfold, etc.). 

As shown in Fig.~\ref{Fig3}b, the angular dependence of the longitudinal and transverse resistance ($R_{xx}$ and $R_{xy}$) of our sample were recorded simultaneously while the magnetic field was rotated within the (111) plane. At T = 2 K, the sixfold (fourfold) harmonic is clearly observed in AMR (PHE) under 8 T field, as can be more evidently seen from the extracted magnitude of each oscillation (Fig.~\ref{Fig3}c). While the fourfold $R_{xx,4}$ and sixfold $R_{xx,6}$ harmonics are well defined in AMR, the sixfold $P_{xy,6}$ term is much suppressed in PHE, referring to the existence of $C_3$ rotation axes along the surface normal based on symmetry analyses \cite{Rout_2017_PRB}. 

Note, the higher harmonics of oscillations diminish practically to zero in the absence of magnetic field (Fig.~\ref{Fig3}b, gray curves) and above T$_\textrm{C}$ (see Supplementary Figure 5 for data at 170 K). Hence, these results confirm the presence of $C_3$ symmetry of the magnetic state, coinciding with the trigonal nature of the underlying lattices \cite{Asaba_2018_PRB}.\\

%chiral-anomaly induced NMR%

\begin{figure}[htp]
\centering
\includegraphics[width=0.95\textwidth]{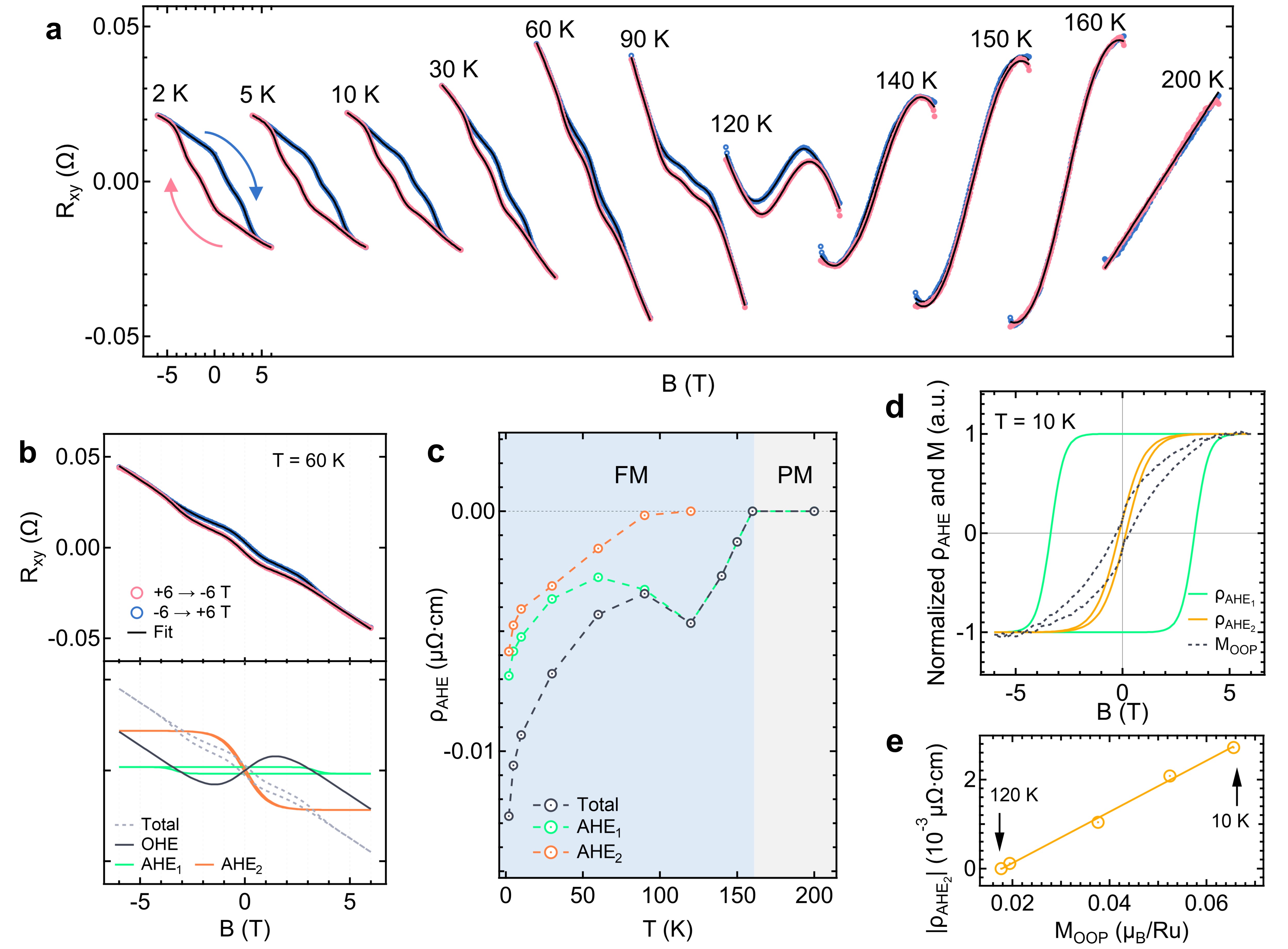}
\caption{\label{Fig4}
{\bf Anomalous Hall effect of trigonal SrRuO$_3$.} \textbf{a,} Evolution of the transverse magnetoresistance R$_{xy}$ from 200 to 2 K. For each curve, the red branch represents the scan from +6 to -6 T field, and vice versa for the blue one. \textbf{b,} A representative fit of data at T = 60 K, using the `two AHEs + nonlinear OHE' model as described in the main text. \textbf{c,} The extracted spontaneous anomalous Hall resistivity $\rho_\textrm{AHE}$ (total signal along with individual components) as a function of temperature. \textbf{d,} Comparison of the normalized hysteresis loop between $\rho_\textrm{AHE}$ of each component and the out-of-plane magnetization $M_\textrm{OOP}$ at 10 K. \textbf{e,} Plot of the linear relationship between the spontaneous $\rho_\textrm{AHE$_2$}$ and $M_\textrm{OOP}$ from 120 to 10 K.}
\end{figure}

{\noindent \bf Ordinary and anomalous Hall effect.}
After demonstrating the structural and magnetic features, we turn to explore the Berry-phase manipulations and plausible topological properties of trigonal SrRuO$_3$ from Hall measurements at a set of temperatures across T$_\textrm{C}$ (Fig.~\ref{Fig4}a). At T = 200 K, the Hall resistance is simple and linear with respect to field. Right below T$_\textrm{C}$, nonlinear and hysteretic $R_{xy}(B)$ curves are captured during 160 - 120 K, as a result of the mixture of AHE with nonlinear ordinary Hall effect (OHE). Strikingly, at lower temperatures from 90 to 2 K, the hysteresis of AHE exhibits a more intricate `two-step' character, indicating the coexistence of two AHE channels with different coercive fields \cite{Kimbell_2020_PRM}. To separate each component from the net Hall signal, the data were fitted using a phenomenological model (see `Methods' for details). A representative case at T = 60 K is displayed in Fig.~\ref{Fig4}b, where the two AHE components and the nonlinear OHE component have been extracted separately.   

The nonlinear OHE can be rationally described using the two-band model, where both electrons and holes contribute to charge transport. This reflects the coexistence of electron and hole pockets on the Fermi surface, plausibly due to the splitting between majority and minority spin bands below the Curie temperature \cite{Hahn_2021_PRL}. The dominant carriers are electrons with low mobility and high concentration ($\mu_{e}$ = 0.14 cm$^2$V$^{-1}$s$^{-1}$, $n_e$ = 7.1$\times$10$^{23}$ cm$^{-3}$ at 2 K). These values are in common with those reported in bulk or (001) SrRuO$_3$ thin films \cite{Shimizu_2014_APL,Majcher_2014_JAP}, indicating the electron pockets from normal quadratic bands. However, it is striking to note the rather high mobility of holes ($\mu_{h}$ = 4.8$\times$10$^3$ cm$^2$V$^{-1}$s$^{-1}$, $n_h$ = 1.1$\times$10$^{16}$ cm$^{-3}$ at 2 K) in our trigonal SrRuO$_3$. In general, existence of highly mobile carriers is rare in complex oxides due to strong correlations, and the observation of such may suggest its topologically nontrivial origin from the band structures \cite{Takiguchi_2020_NC,Fujioka_2019_NC, Veit_2018_NC, Yang_2022_PRB}. Particularly, electrons of high mobility ($\sim$10$^4$ cm$^2$V$^{-1}$s$^{-1}$) have been lately reported in ultra-clean (001) SrRuO$_3$ films, signaling the contributions of Weyl fermions in electrical transport \cite{Takiguchi_2020_NC}. We speculate in a similar fashion that the hole pockets in trigonal SrRuO$_3$ originate from linear Weyl bands \cite{Shekhar_2015_NP}.  

%The dominant carriers are holes with relatively slow mobilities in the high-temperature paramagnetic state. However, A sudden drop of its density $n_h$ and significant rise of its mobility $\mu_h$ takes place while crossing the phase transition, with the occurrence of electron carriers of high density $n_e$ and low mobility $\mu_e$ (bottom panel of Fig.~\ref{Fig3}b). 

We now discuss the AHE. Recap that our trigonal SrRuO$_3$ favors the in-plane magnetic anisotropy parallel to the film surface, observation of AHE would be exotic in the geometry of standard Hall measurements. However, two channels of AHE are clearly probed in trigonal SrRuO$_3$. Fig.~\ref{Fig4}c shows the temperature dependence of the spontaneous Hall resistivity (total $\rho_\textrm{AHE}$ along with individual components $\rho_\textrm{AHE$_1$}$ and $\rho_\textrm{AHE$_2$}$) of trigonal SrRuO$_3$. Essentially, AHE$_1$ is developed right below the Curie temperature, exhibiting non-monotonous evolutions as a function of temperature and significantly broader hysteresis loops than out-of-plane magnetization $M_\textrm{OOP}$ (Fig.~\ref{Fig4}d). These results suggest a momentum-space Berry-phase mechanism for AHE$_1$, where the spontaneous Hall conductivity is intrinsically given by the nonzero Berry curvatures from Weyl pairs near the Fermi energy \cite{Fang_2003_science}. Recall that for the dominant twofold harmonics of AMR (Fig.~\ref{Fig3}b), the resistance at $B \parallel I$ is smaller than $B \perp I$ (i.e. $R_{\parallel} <  R_{\perp}$), at adds with that of a trivial ferromagnetic metal where $R_{\parallel} >  R_{\perp}$ is usually expected as a result of the spin-dependent scattering. This behavior can plausibly be rationalized by the chiral-anomaly induced negative MR effect, in which existence of Weyl nodes of opposite chiralities would lead to additional conducting channels under the geometry of $B \parallel I$ \cite{Takiguchi_2020_NC, Lin_2021_AM}.

On the contrary, AHE$_2$ is only visible below $\sim$120 K and increases monotonously as temperature decreases. The narrow loops of normalized $\rho_\textrm{AHE$_2$}(B)$ and $M_\textrm{OOP}(B)$ resemble each other closely with nearly identical coercive fields (Fig.~\ref{Fig4}d). Especially, the magnitude of spontaneous $\rho_\textrm{AHE$_2$}$ exhibits a linear relationship with respect to $M_\textrm{OOP}$ at zero field (Fig.~\ref{Fig4}e). In this manner, the signal of $\rho_\textrm{AHE$_2$}$ practically mimics the profile of magnetization $M(B, T)$ that is perpendicular to the Hall plane. Hence we tend to believe that AHE$_2$ is of conventional type as commonly expected in an itinerant ferromagnetic system, resulting from field-induced canting and pinning of magnetization along the [111] direction \cite{Nagaosa_RMP_2010}.\\

%The scaling relationship between anomalous Hall conductivity $\sigma_{xy}$ and longitudinal conductivity $\sigma_{xx}$ indicates the R$\bar{3}$c-phase SrRuO$_3$ mainly falls in the intrinsic regime of the unified theory of AHE (Fig.~\ref{Fig3}c), where the origin of AHE is dominated the nonzero Berry curvatures of bands near the Fermi energy combined with the side-jump mechanism \cite{Nagaosa_RMP_2010,Onoda_2008_PRB}. This is consistent with the results from thick (001) SrRuO$_3$ films \cite{Haham_2011_PRB}, whereas extrinsic factors induced by disorders and defects would smear the intrinsic contributions due to dephasing of the Berry phase,  and favor a $\sigma_{xy} \propto \sigma_{xx}^{1.6}$ scaling relationship as recently reported in ultrahtin SrRuO$_3$ films of only several unit cells \cite{Wu_2020_PRB}.\\

\begin{figure}[htp]
\centering
\includegraphics[width=0.98\textwidth]{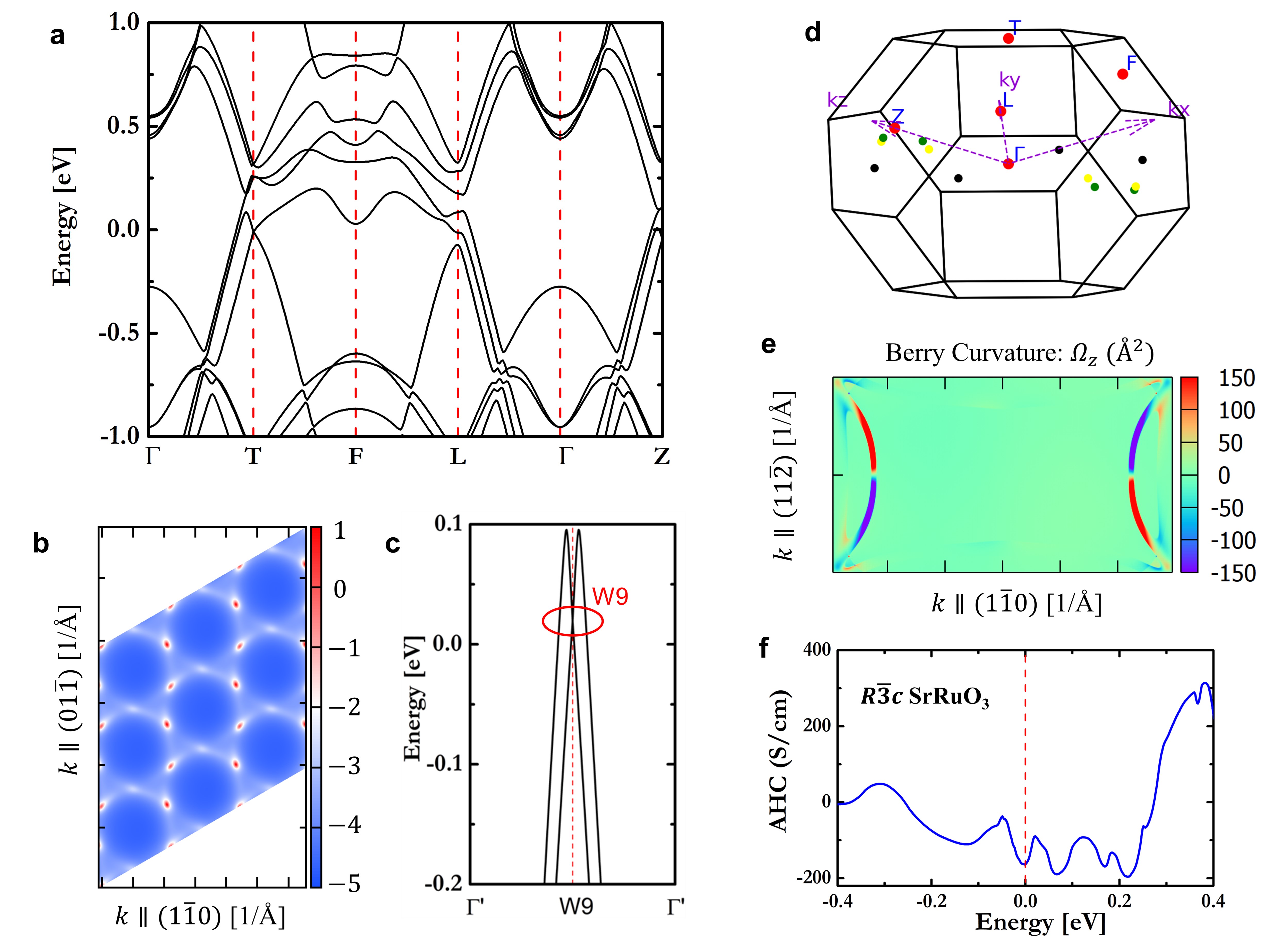}
\caption{\label{Fig5}
{\bf Calculated electronic and topological properties of trigonal SrRuO$_3$ with in-plane magnetization.} 
\textbf{a,} Band structure along the high-symmetry $k$-paths ($\Gamma$: (0 0 0); T: (0.5 0.5 0.5); F: (0.5 0.5 0); L: (0 0.5 0)). 
\textbf{b,} Surface state spectrum (plotted in the logarithmic scale) at the Fermi energy projected on the (111) plane of trigonal SrRuO$_3$. 
\textbf{c,} The enlarged band structure around a linear crossing point denoted as W9, along $\Gamma$' (-0.04452,0.10343,0) to W9 (-0.04452,0.10343,-0.43273).
\textbf{d,} Distribution of the Weyl points over the first Brillouin zone within an energy range $\lvert E_W - E_F \rvert <$ 0.05 eV above and below the Fermi energy. The colors represent Weyl points of different energies (W9 is drawn in green). The Cartesian coordinates ($k_x$, $k_y$, $k_z$) are indicated on the figure and $\Gamma$, T, F, L, and Z denote the high-symmetry momenta. 
\textbf{e,} Mapping of the Berry curvatures in the (111) plane. 
\textbf{f,} Calculated anomalous Hall conductivity (AHC) of trigonal SrRuO$_3$.
}
\end{figure}

{\noindent \bf Electronic band structure calculations.}
Since intrinsic AHE is a direct manifest of the band topology, the observed distinctive behaviors in trigonal SrRuO$_3$ would refer to dramatically manipulated electronic band structures compared to the orthorhombic or tetragonal SrRuO$_3$ phases that are normally stabilized in (001) thin films \cite{Tian_2021_PNAS}. To highlight these information, we performed density functional theory (DFT) calculations to investigate the possible presence and distribution of Weyl points (see `Methods' for more details). 

First, we calculated the energy of SrRuO$_3$ in various space groups. Six most plausible symmetries (cubic $Pm\bar{3}m$; orthorhombic $Pbnm$, $Imma$; trigonal $R\bar{3}c$; monoclinic $C2/c$, $C2/m$) were considered for comparison. To simulate the effect of epitaxial strain, the in-plane lattice parameter of the film was locked to the substrate's value and the out-of-plane lattice parameter was allowed to relax. As shown in Fig.~\ref{Fig1}b, it is notable that the trigonal $R\bar{3}c$ SrRuO$_3$ with $a^-a^-a^-$ rotation pattern exhibits the lowest energy on KTaO$_3$ substrate, rather than the orthorhombic $Pbnm$ with $a^-a^-c^+$ rotation pattern. By calculating the energies of these two phases as a function of (111) strain, we find that the $R\bar{3}c$ phase starts to emerge under small tensile strain, whereas $Pbnm$ is still favored under compressive strain. These results agree well with our experimental observations. 

Next, we calculated the band structure of trigonal SrRuO$_3$ in the non-magnetic case (Supplementary Figure 6). According to the calculated compatibility relations of this band structure, there exist several topological Dirac points around the Fermi level along high-symmetry line $\Gamma$ - $T$. The pattern of surface states at the Fermi level suggests a $C_6$ rotation symmetry due to combined $C_3$ and $C_2$ symmetries (Supplementary Figure 7). By taking the spin-orbit coupling (SOC) into account, we calculated the band structure of the ferromagnetic state with an in-plane spontaneous magnetization as suggested from our experiments. Due to breaking of the time-reversal symmetry, the doubly degenerate Dirac points are expected to split into Weyl pairs. 

Two key features are captured from the calculations. On one hand, along high-symmetry $k$-paths, the Fermi energy crosses conventional quadratic bands, contributing the dominant carriers with low-mobility (Fig.~\ref{Fig5}a). The resultant electron pockets with finite density of states are reflected from the surface state spectrum (Fig.~\ref{Fig5}b). On the other hand, multiple pairs of Weyl nodes are found near the Fermi energy, distributed along low-symmetry $k$-paths in the first Brillouin zone (Supplementary Table 1). The linear dispersion of a representative Weyl point (W9) at an energy $\sim$0.02 eV above the Fermi energy is shown in Fig.~\ref{Fig5}c, and all pairs of Weyl nodes within $\lvert E_W - E_F \rvert <$ 0.05 eV above and below the Fermi energy are plotted in Fig.~\ref{Fig5}d. The high-mobility carriers are likely contributed from these Weyl bands. In addition, Weyl points of opposite chiralities near the Fermi energy serve as source and sink of the Berry curvatures, leading to intrinsic anomalous Hall conductivity (AHC) as illustrated in Fig.~\ref{Fig5}e-f, where the value of AHC depends sensitively on the position of the Fermi energy.

%{\noindent \bf Concluding remarks}\

To conclude, our combined experimental and theoretical investigations demonstrate an emergent trigonal phase of SrRuO$_3$ induced by (111) tensile strain, and reveal its peculiar magnetic and topological features which are remarkably different from bulk. These results provide ubiquitous insights into the scenario of film-substrate interplay for perovskite's (111) heteroepitaxy. Moreover, it also highlights the opportunities for realizing intriguing correlated and topological quantum states of matter in oxides, especially for quantum anomalous Hall effect and topological insulator at the quasi-two-dimensional limit.

%In addition, a cover letter needs to be written with the
%following:
%\begin{enumerate}
 %\item A 100 word or less summary indicating on scientific grounds
%why the paper should be considered for a wide-ranging journal like
%\textsl{Nature} instead of a more narrowly focussed journal.
 %\item A 100 word or less summary aimed at a non-scientific audience,
%written at the level of a national newspaper.  It may be used for
%\textsl{Nature}'s press release or other general publicity.
 %\item The cover letter should state clearly what is included as the
%submission, including number of figures, supporting manuscripts
%and any Supplementary Information (specifying number of items and
%format).
 %\item The cover letter should also state the number of
%words of text in the paper; the number of figures and parts of
%figures (for example, 4 figures, comprising 16 separate panels in
%total); a rough estimate of the desired final size of figures in
%terms of number of pages; and a full current postal address,
%telephone and fax numbers, and current e-mail address.
%\end{enumerate}

\newpage

\begin{methods}

\noindent{\bf {Sample fabrication.}}\ The (111) oriented SrRuO$_3$ thin films were grown on 5$\times$5 mm$^2$ (111) KTaO$_3$ single crystalline substrates by pulsed laser deposition. SrRuO$_3$ ceramic target was ablated using a KrF excimer laser ($\lambda$ = 248 nm, energy density $\sim$2 J/cm$^{2}$) with a repetition rate of 2 Hz. The deposition was carried out at a substrate temperature of 730 $^{\circ}$C, under an oxygen atmosphere of 75 mTorr. The films were post-annealed at the growth condition for 15 min, and then cooled down to room temperature. The whole deposition process was {\it{in-situ}} monitored by reflective-high-energy-electron diffraction.

\noindent{\bf {Scanning transmission electron microscopy.}}\ The scanning transmission electron microscopy (STEM) measurements were carried out using a double spherical aberration-corrected JEM-ARM200F, operated at 200 kV. The sample was cut and projected onto the (11$\bar{2}$)$_\textrm{pc}$ plane. The high-angle annular dark-field (HAADF) imaging was taken using the collection semi-angle of about 70-250 mrad. Positions of the atoms were extracted using the CalAtom software, which basically normalized the brightness of the image and calculated in the vicinity of each atom the local maximum as the round center. This process gave rise to a simulated matrix where the relative position of each atom is labeled with two coordinates, based on which the bucking of the Sr and O sublattices were analyzed.

\noindent{\bf {Magnetization measurements.}}\ The magnetization measurements were carried out using a Magnetic Property Measurement System (MPMS-3, Quantum Design). After zero-field cooling the sample down to 10 K, the temperature dependence of magnetization was measured from 10 K to 300 K in an external field of 1000 Oe applied in the film plane. Field dependence of magnetization was measured at 10 K by applying the magnetic field along [111], [1$\bar{1}$0], and [11$\bar{2}$] direction, respectively.

\noindent{\bf {Transport measurements.}}\ The electrical transport measurements were performed in a Physical Property Measurement System (PPMS, Quantum Design) with the 4-point contact method. During the magneto-transport experiments, the magnetic field was applied along the [111] direction, and the current was driven along the [1$\bar{1}$0] direction. At each temperature, the longitudinal and transverse resistance were recorded while sweeping the field in a cycle from +6 T to -6 T to +6 T. Standard symmetrization (anti-symmetrization) treatment was further applied using these two branches of data on the longitudinal (transverse) resistance to achieve pure signals of the magnetoresistance (Hall resistance).

\noindent{\bf {Details of analyses on anomalous Hall effect.}}\ At temperatures below the magnetic phase transition, the overall Hall effect of SrRuO$_3$ on KTaO$_3$ is attributed from both the ordinary Hall effect (OHE) and the anomalous Hall effect (AHE), $R_{xy}(B) = R_{xy}^\textrm{OHE}(B) + R_{xy}^\textrm{AHE}(B)$. In particular, at lower temperatures (e.g. T = 60 K as shown Fig.~\ref{Fig3}a), the OHE component is nonlinear and the AHE component contains two contributions with different coercive fields. Thus we introduce the phenomenological model `two AHEs + nonlinear OHE' to describe the Hall data. The AHE component $R_{xy}^\textrm{AHE}(B)$ is expressed as:
\begin{align}
R_{xy}^\textrm{AHE}(B) = R_{xy}^\textrm{AHE1}(B) + R_{xy}^\textrm{AHE2}(B) = R_1^\textrm{AHE}tanh[\omega_1(B - B_{c,1})] + R_2^\textrm{AHE}tanh[\omega_2(B - B_{c,2})]
\end{align}
where $R_i^\textrm{AHE}$ and $B_{c,i}$ represent the spontaneous AH resistance and the coercive field of hysteresis loop, respectively; $\omega_i$ is a parameter describing the slope of loop at $B_{c,i}$.
The nonlinear OHE component $R_{xy}^\textrm{OHE}(B)$ is expressed based on the conventional two-carrier model:
\begin{align}
R_{xy}^\textrm{OHE}(B) = \frac{B}{d \cdot e} \frac{(n_h \mu_h^2 - n_e\mu_e^2) + (n_h - n_e)\mu_h^2 \mu_e^2 B^2}{(n_h\mu_h + n_e\mu_e)^2 + (n_h - n_e)\mu_h^2 \mu_e^2 B^2}
\end{align}
where $n_h$ ($n_e$), $\mu_h$ ($\mu_e$), $d$ and $e$ represent the density of holes (electrons), the mobility of holes (electrons), the thickness of films, and the elementary charge, respectively.\

To extract the parameters ($n_h$, $n_e$, $\mu_h$, $\mu_e$) from the fit, we denote $\alpha = \frac{\mu_e}{\mu_h}$, $\beta = \frac{n_e}{n_h}$ ($\alpha$, $\beta$ $>$ 0). Then the $R_{xy}^\textrm{OHE}(B)$ can be rewritten as:
\begin{align}
R_{xy}^\textrm{OHE}(B) = \frac{B}{d \cdot e \cdot n_h} \frac{1}{1-\beta} \frac{\frac{(1-\alpha^2\beta)(1-\beta)}{(1+\alpha\beta)^2} + \frac{(1-\beta)^2\alpha^2}{(1+\alpha\beta)^2}\mu_h^2B^2}{1+\frac{(1-\beta)^2\alpha^2}{(1+\alpha\beta)^2}\mu_h^2 B^2}
\end{align}
In the limits of $B \to \infty$ and $B \to 0$, $R_{xy}^\textrm{OHE}(B \to \infty) = \frac{B}{d \cdot e \cdot n_h} \frac{1}{1-\beta}$; $R_{xy}^\textrm{OHE}(B \to 0) = \frac{B}{d \cdot e \cdot n_h} \frac{1-\alpha^2\beta}{(1+\alpha\beta)^2}$.
We further denote:
\begin{align}
&K = \frac{1}{d \cdot e \cdot n_h} \frac{1}{1-\beta};\\
&C = \frac{(1-\alpha^2\beta)(1-\beta)}{(1+\alpha\beta)^2};\\
&D = \frac{(1-\beta)^2\alpha^2}{(1+\alpha\beta)^2}\mu_h^2
\end{align}
where $K$ and $C$ can be either positive or negative; $D \geq 0$. In this manner, the OHE component is eventually expressed as:
\begin{align}
R_{xy}^\textrm{OHE}(B) = K \cdot B \cdot \frac{C + D \cdot B^2}{1 + D \cdot B^2}
\end{align}
Meanwhile, based on the two-carrier model, the longitudinal resistance is expressed as:
\begin{align}
R_{xx}(B) = \frac{1}{d \cdot e} \frac{(n_h \mu_h + n_e\mu_e)(1 + \mu_h\mu_e B^2)}{(n_h\mu_h + n_e\mu_e)^2 + \mu_h^2\mu_e^2(n_e - n_h)^2 B^2}
\end{align}
At $B = 0$, it is rewritten as:
\begin{align}
\lambda = R_{xx}(B = 0) = \frac{1}{d \cdot e} \frac{1}{n_h\mu_h + n_e\mu_e} = \frac{1}{d \cdot e} \frac{1}{n_h\mu_h} \frac{1}{1+\alpha\beta}
\end{align}
Here, $\lambda$ can be directly obtained from the measured longitudinal resistance at zero field.

In this end, the measured longitudinal and transverse magnetoresistance at individual temperatures were simultaneously fit using equation (3), (9) and (11), from which one can achieve the fitting parameters \{$R_i^\textrm{AHE}$, $\omega_i$, $B_{c,i}$\} relevant to AHE and those \{$K$, $C$, $D$, $\lambda$\} relevant to OHE. Eventually, the carrier densities and mobilities $n_h$, $n_e$, $\mu_h$, $\mu_e$ can be achieved by solving the combined four equations of (6), (7), (8), (11) with a simple Mathematica program.

\noindent{\bf {Anisotropic magnetoresistance and planar Hall effect.}}\ For the in-plane angle-dependent transport experiments, the current $I$ was applied along the [1$\bar{1}$0] direction. The magnetic field $B$ lied within the (111) plane, and rotated with an angle $\theta$ relative to the current direction. The measurements were started with an offset of $\theta$ $\sim$67 $^{\circ}$, and the longitudinal resistance $R_{xx}$ and the transverse resistance $R_{xy}$ were measured simultaneously as a function of the angle. The obtained experimental data on $R_{xx}$ and $R_{xy}$ were fit using the following expansions up to the eightfold harmonics:
\begin{align}
&R_{xx} = \displaystyle\sum_{n=0}^{4} R_{xx,2n}  = R_{0} + \displaystyle\sum_{n=1}^{4} R_{2n}  cos(2n(\theta + \theta_{2n})) \\
&R_{xy} = \displaystyle\sum_{n=0}^{4} P_{xy,2n} = P_{0} + \displaystyle\sum_{n=1}^{4} P_{2n}  sin(2n(\theta + \theta_{2n}))
\end{align}
where $R_{2n}$ ($P_{2n}$) is the magnitude of each harmonics in $R_{xx}$ ($R_{xy}$). Note, due to the slight misalignment of contacts, signals from the transverse resistance could be mixed with small portions of the longitudinal resistance, giving rise to the $P_0$ term in $R_{xy}$. The corresponding anisotropic magnetoresistance (AMR) and planar Hall effect (PHE) are defined as $\textrm{AMR} = (R_{xx} - R_0)/R_0 \times 100\%$; $\textrm{PHE} = R_{xy} - P_0$, respectively.

\noindent{\bf {Theoretical calculations.}}\ 
We performed first principles calculations on the electronic properties of magnetic material SrRuO$_3$ utilizing
the generalized gradient approximation (GGA) with the revised Perdew-Burke-Ernzerhof for solids (PBEsol)\cite{PS}
and the projector augmented wave method\cite{PAW} as implemented in the VASP\cite{VASP}.
We constructed equivalent [111] crystalline direction of cubic for orthorhombic (space group:62) and trigonal (space group:167) lattice in order to match the KTO substrate and further tune the strain using fixed relaxed process.
Under this substrate effect, the $R\bar{3}c$ lattice with the point group $D_{3d}$ (-3m) takes up the ground state, which is transformed into primitive cell for following calculations.
Note, the $e_g$ and $t_{2g}$ orbitals under $O_h$ point group from cubic symmetry viewpoint has been changed into $e_g$ and $a_{1g}$ + $e_g$ orbitals under $D_{3d}$, respectively. Meantime, basis function of the new $e_g$ orbital is ($x^2 - y^2$, $xy$) or ($xz$, $yz$) and the corresponding function of $a_{1g}$ orbital is only $z^2$. The generators of this parent structure include identity, $C_3$, \{$C_2$, (0, 0, 1/2)\}, inversion, \{1, \{2/3, 1/3, 1/3\}\} pure translation group elements. Therefore, without considering the magnetization of Ru atoms, the non-magnetic band structure will keep the Kramer’ double degenerate states due to protection by time-reversal symmetry and space inversion symmetry (Supplementary Figure 6).
The $\Gamma$-center $k$-mesh was set as 7$\times$7$\times$7 for integration over Brillouin zone and kinetic energy cutoff was equal to 500 eV. The spin-orbital coupling was self-consistently included in the electronic computations.
Considering the correlated effect of Ru-d orbital, Hubbard U value\cite{DFT+U} was set as 2 eV, which gives rise to 2 $\mu_{B}$ localization magnetic moment. Symmetry analysis based on magnetic topological quantum chemistry theory have been studied together with compatibility relations\cite{MTQC,CR,wang_mb}, which suggests enforced semimetal feature. Meantime, tight-binding model was also constructed by projecting Ru-d as well as O-p orbital into localized wannier basis\cite{wannier}. Chirality of nodes and anomalous hall effect were calculated using wanniertools software\cite{wt}.

\noindent{\bf {Data availability.}}\ All relevant data are available from the authors upon reasonable request.

\noindent{\bf {Code availability.}}\ All relevant codes are available from the authors upon reasonable request.

\end{methods}

\begin{addendum}
 \item [Acknowledgements.]
%The authors deeply acknowledge T. Ying, J. Sun, J. Zhao for numerous insightful discussions. 
This work is supported by the National Natural Science Foundation of China under Grant No. 12204521, the National Key R\&D Program of China (No. 2017YFA0303600), and the Strategic Priority Research Program (B) of the Chinese Academy of Sciences (No. XDB33000000). Q.Z. acknowledges the support from Beijing Natural Science Foundation (Z190010) and National Natural Science Foundation of China (52072400, 52025025).  

 \item [Author contributions.]
X.L. and J.G. conceived and designed the projects. Z.D., Z.W., F.Y. and X.L. fabricated the samples. Z.D. and Z.W. performed the transport and magnetization measurements, and analyzed the data. Q.Z. and L.G. carried out the scanning transmission electron microscopy imaging. %J.B. and Y.C. did the reciprocal space mapping measurement. 
X.C. and Z.Z. performed the {\it{ab-initio}} theoretical calculations. Z.D., X.C., X.L. and J.G. wrote the manuscript. All authors contributed to the discussions and commented on the manuscript. 

 \item [Additional information.]
Supplementary information is available in the online version of this manuscript. Reprints and permissions information is available online at www.nature.com/reprints. 
$\dagger$ These authors contributed equally to this work. 
%Correspondence and requests for materials should be addressed to X.L. (xiaoran.liu@iphy.ac.cn); Z.Z. (zhong@nimte.ac.cn); J.G. (jdguo@iphy.ac.cn).

 \item [Competing interests.]
The Authors declare no Competing Financial or Non-Financial Interests.
\end{addendum}

%\subsection{Method subsection.}
%Here is a description of a specific method used.  Note that the
%subsection heading ends with a full stop (period) and that the
%command is \verb|\subsection{}| not \verb|\subsection*{}|.

%% Put the bibliography here, most people will use BiBTeX in
%% which case the environment below should be replaced with
%% the \bibliography{} command.

%% Here is the endmatter stuff: Supplementary Info, etc.
%% Use \item's to separate, default label is "Acknowledgements"

%%
%% TABLES
%%
%% If there are any tables, put them here.
%%

%\begin{thebibliography}{99}

%\bibitem{Balents_ARCMP_2014}
%Witczak-Krempa, W., Chen, G., Kim, Y. \& Balents, L.~Correlated quantum phenomena in the strong spin-orbit regime. {\it{Annu. Rev. Condens. Matter Phys.}} \textbf{5}, 57-82 (2014).

%\end{thebibliography}

\end{document}